\documentclass[journal=jpcbfk,manuscript=article]{achemso}

\usepackage{graphicx,xcolor}
\usepackage{amsmath, bbm}
\usepackage{mathtools}
\usepackage{braket}
\usepackage{color}
\usepackage{siunitx}

\title{Quantum Efficiency of Single Dibenzoterrylene Molecules in \textit{para}-Dichlorobenzene at Cryogenic Temperatures}
\author{Mohammad Musavinezhad}
\affiliation{Max Planck Institute for the Science of Light, D-91058 Erlangen, Germany}
\alsoaffiliation{Department of Physics, Friedrich Alexander University Erlangen-Nuremberg, D-91058 Erlangen, Germany}
\author{Alexey Shkarin}
\author{Dominik Rattenbacher}
\author{Jan Renger}
\author{Tobias Utikal}
\affiliation{Max Planck Institute for the Science of Light, D-91058 Erlangen, Germany}
\author{Stephan G\"otzinger}
\affiliation{Department of Physics, Friedrich Alexander University Erlangen-Nuremberg, D-91058 Erlangen, Germany}
\alsoaffiliation{Max Planck Institute for the Science of Light, D-91058 Erlangen, Germany}
\alsoaffiliation{Graduate School in Advanced Optical Technologies (SAOT), Friedrich Alexander
University Erlangen-Nuremberg, D-91052 Erlangen, Germany}
\author{Vahid Sandoghdar}
\affiliation{Max Planck Institute for the Science of Light, D-91058 Erlangen, Germany}
\alsoaffiliation{Department of Physics, Friedrich Alexander University Erlangen-Nuremberg, D-91058 Erlangen, Germany}
\email{vahid.sandoghdar@mpl.mpg.de}

\begin{document}

\begin{abstract}
We measure the quantum efficiency (QE) of individual dibenzoterrylene (DBT) molecules embedded in \textit{para}-dichlorobenzene at cryogenic temperatures. To achieve this, we apply two distinct methods based on the maximal photon emission and on the power required to saturate the zero-phonon line. We find that the outcome of the two approaches are in good agreement, reporting a large fraction of molecules with QE values above \SI{50}{\%}, with some exceeding \SI{70}{\%}. Furthermore, we observe no correlation between the observed lower bound on the QE and the lifetime of the molecule, suggesting that most of the molecules have a QE exceeding the established lower bound. This confirms the suitability of DBT for quantum optics experiments. In light of previous reports of low QE values at ambient conditions, our results hint at the possibility of a strong temperature dependence of the QE.
\end{abstract}

\maketitle

\section{Introduction}
Photoluminescence quantum yield (QY) is a key property of an optical emitter, as it determines the emitter's efficiency of converting the incoming light to luminescence. This quantity plays a crucial role in a variety of applications such as bioimaging\cite{Lakowicz2006principles} or lasing\cite{Siegman1986lasers}. QY is typically defined as the ratio of the number of emitted to absorbed photons, and therefore depends on both the emission and absorption properties of the emitter under study. 

Given its technological and fundamental importance, there exists an extensive body of work investigating QY in a variety of systems\cite{Johansen2008,Olutas2015,Wurth2012,Cranfill2016}. The great majority of such studies are, however, conducted via ensemble measurements, where a macroscopic quantity of photoluminescent material is typically illuminated with a light of known characteristics. Carefully calibrated measurements of the emission (e.g., using a sample of known QY) and absorption (usually determined via thermal effects) are used to extract the QY \cite{Wurth2015}. In addition to the challenge of performing accurate calibrations, the very nature of this approach makes it insensitive to inter-emitter variations that are inherent to the specific emitter type \cite{Orfield2016,Mohtashami2013} or arise from differences in their local environment, especially in the solid state \cite{Chizhik2011,Chu2017a,Kwadrin2012}. As a result, our quantitative and first-principle understanding of the QY remains incomplete.

The recent progress of nano-optics has invoked the use of single quantum emitters in a variety of applications, ranging from biological super-resolution microscopy to quantum information processing \cite{Sandoghdar-NL2020}. The photophysics of the emitter and therefore its QY play a central role in nearly all these applications. However, single-molecule QY measurements are very challenging because they require accurate measurements of weak optical powers and minute thermal dissipations. In an alternative approach, one  compares the radiated photon rate of a single emitter to the total decay rate of its excited state and defines the quantum efficiency (QE) as 
\begin{equation}
\rm{QE}=\gamma_\mathrm{r}/\gamma_\mathrm{tot} \,\, ;\,\, \gamma_\mathrm{tot}=\gamma_\mathrm{r}+\gamma_\mathrm{nr}
\end{equation} 
where $\gamma_\mathrm{r}$, $\gamma_\mathrm{nr}$ and $\gamma_\mathrm{tot}$ denote the radiative, nonradiative, and total decay rates of the given quantum state, respectively. Hence, while QY is a more technologically relevant quantity, QE establishes a fundamental emitter property that can be more readily used in different excitation schemes.

Various methods have been used for measuring QE of single molecules. In one class of experiments\cite{Chu2017a}, the collection and detection efficiencies of the measurement setup are carefully calibrated, so that the detected photon rate (power, $P$) can be directly related to the radiative decay rate $\gamma_\mathrm{r}$ of the emitter through the relation $P=\hbar\omega\gamma_\mathrm{r}\rho_{ee}$, where $\rho_{ee}$ is the excited state population and $\hbar$ is the Planck's constant. In addition, the total decay rate $\gamma_\mathrm{tot}$ is directly assessed by measuring the lifetime of the excited state.

Another line of studies is inspired by the pioneering work  of K. Drexhage on the modification of the fluorescence lifetime when an emitter is placed close to an interface\cite{Drexhage1974}. Here, one exploits the fact that changes to the local electromagnetic environment, e.g., the refractive index of the surrounding\cite{Brokmann2004}, a movable mirror\cite{Buchler2005} or a tunable optical cavity\cite{Chizhik2011} modify $\gamma_\mathrm{r}$ but leave $\gamma_\mathrm{nr}$ unchanged. Since this method only requires lifetime measurements, it circumvents the difficulties associated with the calibration of excitation and emission efficiencies. The downside of the approach is, however, its strong sensitivity to the exact position and orientation of the emitter with respect to the physical boundaries. 

In this article, we perform QE measurements on individual dibenzoterrylene (DBT) molecules embedded in an organic crystal (\textit{para}-dichlorobenze, \textit{p}DCB) at $T=2$\,K \cite{Jelezko1996}. DBT belongs to the family of polycyclic aromatic hydrocarbons (PAH) which has been used in a number of quantum optical studies because of their high spectral stability, strong zero-phonon lines (ZPL) and negligible dephasing when they are embedded in a suitable matrix at low temperatures\cite{Toninelli2021}. The results of the experiments have been consistent with high QE values \cite{Wang2019,Pscherer2021,Trebbia2022,Wrigge2008}although quantitative QE studies of these systems have been rare \cite{DeVries1979}. In fact, explicit reports are missing at the single emitter level. Concrete QE reports of these systems have been based on ensemble \cite{Kwadrin2012,Erker2022} or single-molecule \cite{Chu2017a,Buchler2005,Erker2022} measurements at room temperature. Interestingly, a recent publication has reported QE values of 35\% and below for DBT at room temperature\cite{Erker2022}. The dependence of the fluorescence lifetime and QE on the $S_1$-$S_0$ transition energy was interpreted in light of the energy gap law (EGL) for the nonradiative decay\cite{Englman1970,Turro2010,Jang2021}, which predicts that the internal conversion (IC) rate should grow exponentially as the transition wavelength becomes longer. Due to the near-infrared transition of DBT (between \SI{700}{nm} and \SI{800}{nm}, depending on the matrix)\cite{Jelezko1996,Hofmann2005,Verhart2016}, EGL is expected to enhance IC and, thus, $\gamma_{\rm nr}$\cite{Kwadrin2012,Nicolet2007,Turro2010}. Our study aims to clarify whether the room-temperature reports for DBT also hold for its cryogenic applications.

\section{Experimental Methods}

\begin{figure}[t] 
\includegraphics[width=0.5\textwidth]{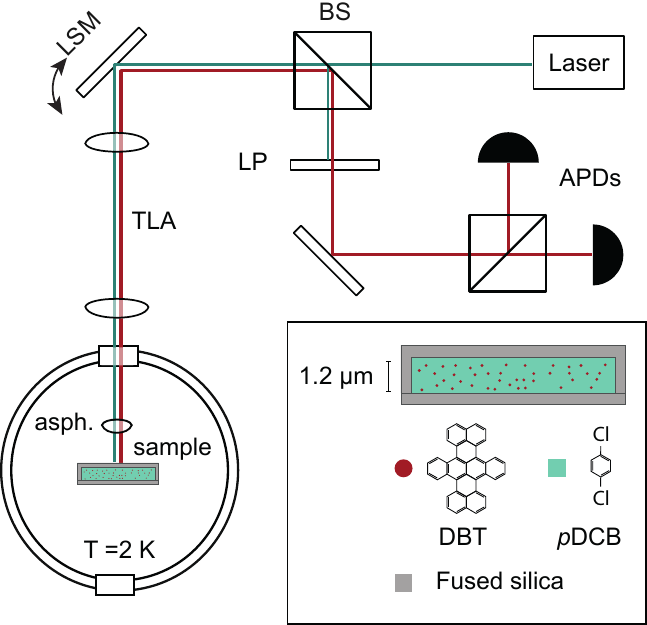}
\centering
\caption{Optical setup and the sample schematics. BS: beam-splitter; LSM: laser scanning mirror; TLA: telecentric lens assembly; asph: aspherical lens; LP: long-pass fluorescence filter; APD: avalanche photodiode. The inset shows the cross-section of the fused silica channel filled with DBT-doped \textit{p}DCB.}
\label{fig:setup} 
\end{figure}

The sample preparation is similar to our earlier work\cite{Rattenbacher2019}. In short, we first prepare a solution of DBT in \textit{p}DCB at a concentration of about \SI{100}{ppm}. This solution is then melted (\SI{53}{\degree C} melting temperature) on a hot plate and introduced into a \SI{1.2}{\micro m}-thick channel formed between two silica substrates. After crystallization, we perform partial re-melting with a slower crystallization step, which produces homogeneous \textit{p}DCB crystals with roughly \SI{100}{\micro m} lateral size. Finally, this sample is inserted into the cryostat and cooled below \SI{2}{K}.

The basis of the experimental setup is a home-built cryogenic confocal microscope, which allows for single-molecule imaging (see Fig.\,\ref{fig:setup}). To address individual molecules, we excite their narrow lifetime-limited zero-phonon-lines (ZPL, typical linewidth of \SI{25}{MHz}) using a narrow-band ($<\SI{1}{MHz}$) continuous-wave Ti:Sapphire laser. The laser polarization can be controlled to ensure the best matching to the molecule's in-plane dipole, and its power can be adjusted using a set of neutral density filters. The coarse sample alignment and the optical focus adjustment are achieved using cryogenic nanopositioners, while the fine positioning of the laser beam can be realized using a laser scanning mirror (LSM) in combination with a telecentric 4\textit{f} lens assembly. The light emitted by the molecule is collected in reflection and sent through a tunable fluorescence filter that blocks the excitation laser  but passes the red-shifted fluorescence. This light is then guided onto a pair of avalanche photodiodes (APDs) arranged in a Hanbury-Brown and Twiss configuration. This assembly allows us to determine the molecular excited state lifetime and inter-system crossing (ISC) rate via photon autocorrelation measurements.

\section{Results and Discussion}

We measured the properties of 44 molecules at various locations within the sample with transition wavelengths between \SI{743.6}{nm} and \SI{745}{nm}. In order to obtain a fair representation of the overall distribution, we made sure to avoid any selection based on apparent brightness or linewidth. The only exclusion criterion was the vicinity to cracks in the matrix so as to avoid distorted excitation or emission patterns.

\begin{figure}[t!] 
\includegraphics[width=0.5\textwidth]{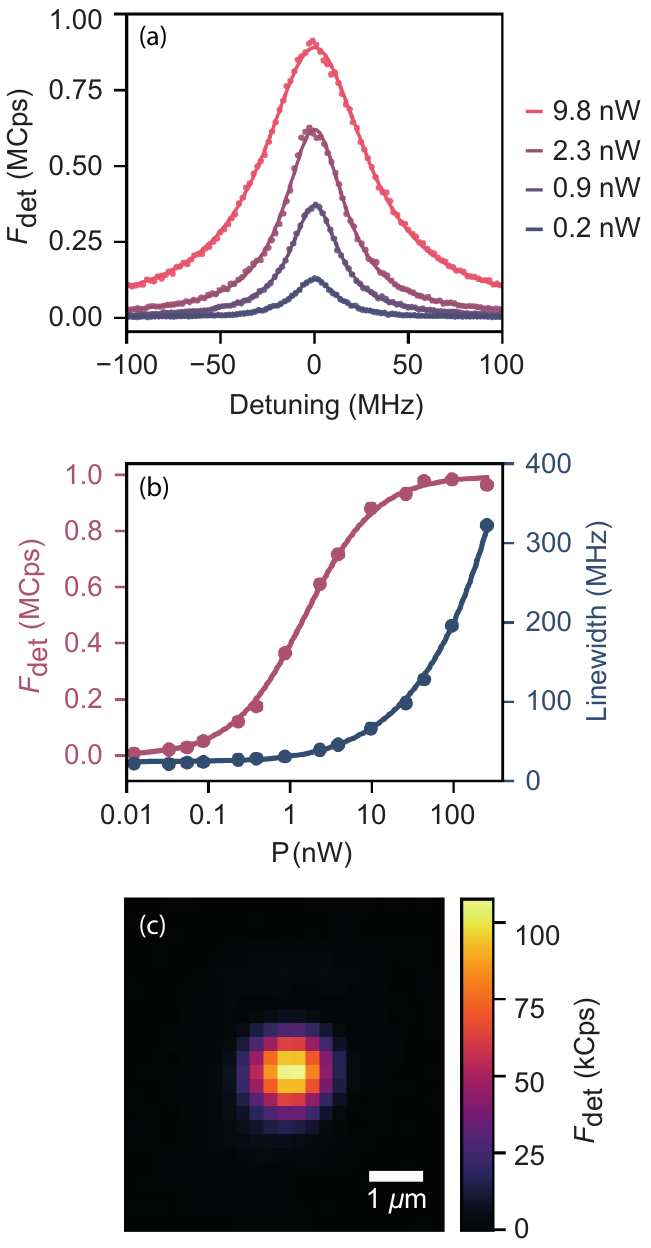}
\centering
\caption{(a) Detected fluorescence counts $F_\mathrm{det}$ of a single molecule as a function of laser-ZPL detuning for varying optical power. The molecule ZPL wavelength is $\lambda=\SI{743.7}{nm}$. (b) Fluorescence count rate (red) and linewidth (blue) plotted against the excitation power. The power is corrected for the setup excitation efficiency $\eta_\mathrm{exc}$, and the fluorescence data is corrected for the APD saturation caused by its dead time. The black lines are fits to the data according to a semi-classical theory yielding $F_\mathrm{det}(\infty)=\SI{1.00}{MCps}$ and $P_\mathrm{sat}=\SI{1.6}{\nano\watt}$. (c) Fluorescence excitation point spread function (PSF) of the same molecule. The power was set to $P=0.05P_\mathrm{sat}\ll P_\mathrm{sat}$ to avoid PSF distortion. The extracted effective illumination area is $A_\mathrm{eff}=\SI{2.4}{\micro\meter\squared}$.} 
\label{fig:Sat_Psf} 
\end{figure}

For each selected molecule, we optimized the incident light polarization and the focus position to achieve the best excitation efficiency and performed a series of laser frequency scans for varying incident powers. As can be seen in Fig.\,\ref{fig:Sat_Psf}(a), such measurements produce a series of Lorentzians with power-dependent height and linewidth. This dependence is plotted in Fig.\,\ref{fig:Sat_Psf}(b) and follows the well-known saturation profile, where the linewidth grows with increasing powers while the emission maximum saturates to a constant value. The results are well described by the standard semi-classical theory for two-level emitters\cite{Walls_milburn_2008, Scully, Tannoudji}, from which we extract two parameters: the detected fluorescence count rate $F_\mathrm{det}(\infty)$ at high excitation power and the saturation power $P_\mathrm{sat}$ at the molecule position. Both parameters characterize the strength of the emitter-light interaction and can, therefore, be used to extract $\gamma_\mathrm{r}$. In addition, we can measure $\gamma_\mathrm{tot}$ either via the excited state lifetime, or directly as the low-power spectroscopic linewidth of the ZPL under the assumption of negligible dephasing. We, thus, determine the QE of single molecules according to two independent measurement methods.

The first approach is based on analyzing $F_\mathrm{det}(\infty)$. The semi-classical theory predicts that in our resonant driving scheme at high saturation, $\rho_{ee}=1/2$, leading to photon emission rate $F(\infty)=\gamma_\mathrm{r}/2$. To relate the detected APD counts $F_\mathrm{det}$ to the fluorescence rate, we analyzed the total detection efficiency of our setup $\eta_\mathrm{tot}=\eta_\mathrm{coll}\eta_\mathrm{tr}\eta_\mathrm{det}\eta_\mathrm{spec}$. Here, $\eta_\mathrm{coll}$ is the collection efficiency of the aspherical lens, $\eta_\mathrm{tr}$ accounts for the transmission losses of the various optical elements in the detection path, $\eta_\mathrm{det}$ is the APD detection efficiency at 744\,nm, and $\eta_\mathrm{spec}$ is the effective detected fraction of the emission, including spectral dependence of the detection efficiency and filtering. We extract the above-mentioned efficiencies independently and find $\eta_\mathrm{tr}=\SI{69}{\%}$ and $\eta_\mathrm{det}=\SI{55}{\%}$ based on the setup and the APD calibrations. The effective collected spectrum fraction $\eta_\mathrm{spec}$ is more difficult to determine because it depends on the chromatic aberrations of the setup (most notably, the aspherical lens), the transmission characteristics of the fluorescence filter, and the drop of the APD detection efficiency with longer wavelengths. Since we need to filter out the excitation laser at the ZPL frequency, the upper limit for $\eta_\mathrm{spec}$ in our detection scheme is $(1-\alpha)$, where $\alpha$ is the emission branching ratio given by the fraction of the emission contained in the ZPL to the total fluorescence, including the red-shifted emission. In our analysis, we keep $\eta_\mathrm{spec}$ as an adjustable parameter, which is taken to be the same for all measured molecules.
The final parameter, $\eta_\mathrm{coll}$, is the least precisely known. Due to the anisotropic dipole emission pattern, $\eta_\mathrm{coll}$ is very strongly dependent on the out-of-plane dipole angle $\theta$, varying between $\eta_\mathrm{coll}^\mathrm{H}=\SI{9}{\%}$ for a horizontally-oriented dipole ($\theta=0$) and $\eta_\mathrm{coll}^\mathrm{V}=\SI{1}{\%}$ for a vertically-oriented dipole ($\theta=\pi/2$). In general, $\eta_\mathrm{coll}$ can be expressed as a weighted combination $\eta_\mathrm{coll}(\theta)=\eta_\mathrm{coll}^\mathrm{H}\cos^2(\theta)+\eta_\mathrm{coll}^\mathrm{V}\sin^2(\theta)$. Unfortunately, the preparation method of our sample does not produce predefined molecule orientations\cite{Toninelli2010,Pfab2004}. Moreover, the relatively low numerical aperture of our collection lens (0.77) does not let us estimate $\theta$ \cite{Lieb2004}. As a result, the measured QE values can vary by almost an order of magnitude depending on the orientation of a molecule. Hence, this measurement scheme only provides a lower bound for the QE.

The second QE measurement method relies on the saturation power $P_\mathrm{sat}$. It can be shown, the $P_\mathrm{sat}$ is related to the radiative decay rate $\gamma_\mathrm{zpl}=\alpha\gamma_{\rm r}$ that takes place via ZPL as
\begin{equation}
    P_\mathrm{sat}\cos^2(\theta)=\frac{\pi}{3}
    \frac{n^2}{\lambda^2} 
    A_\mathrm{eff}\hbar\omega_{\rm zpl}
    \frac{\gamma_\mathrm{tot}^2}{\gamma_\mathrm{zpl}}
    \label{eq:P_sat},
\end{equation}
where $n$ is the refractive index of the host matrix, $\lambda$ is the vacuum wavelength of the ZPL, $\omega_{\rm zpl}$ is the corresponding angular transition frequency, and $A_\mathrm{eff}$ is the effective area of the excitation beam. Given this equation and the knowledge of the branching ratio $\alpha$, we can calculate $\gamma_\mathrm{zpl}$ from $P_\mathrm{sat}$ and then relate it to the total radiative decay rate $\gamma_\mathrm{r}$. As in the previous method, we can extract $\gamma_\mathrm{tot}$ from the spectroscopic linewidth at low excitation powers. The transition frequency is readily obtained from high-resolution fluorescence scans. The parameter $n$ is more difficult to estimate due to the birefringence of the host material and the uncertainty in the reference data\cite{Manghi1967}; in our calculations we have assumed $n=1.6$.

To measure the effective mode area $A_\mathrm{eff}$, we used the molecule itself as a local intensity probe. Here, we employed the LSM to raster scan the position of the resonant focused beam over the molecule while recording its fluorescence. As long as the laser power is significantly below saturation, the fluorescence is proportional to the local optical intensity, and the measured data directly provide the intensity distribution $I(\mathbf{r})$. An example of a resulting map is shown in Fig.\,\ref{fig:Sat_Psf}(c). Next, we evaluate $A_\mathrm{eff}=\int I(\mathbf{r})\mathrm{d}A/I(\mathbf{r}_\mathrm{mol})$, where $\mathbf{r}_\mathrm{mol}$ is the molecule's position. To ensure that the focused light is transversely polarized and that there were no clipping losses at the aspherical lens, we reduced the diameter of the incident laser beam. We note that the main source of uncertainty in this method is again the angle $\theta$ between the molecular dipole and the substrate plane. However, since $\theta$ affects the QE measurements following the same trend in both methods, it still lets us put an estimate on other sources of uncertainty, which are mostly different between the two methods.

Before we present the experimental data, we mention that we have neglected several effects in the analysis described above. First, we do not include the contribution of the ISC\cite{Bernard1993,Basche2008}, which involves long-lived shelving states and is power dependent. However, our intensity autocorrelation measurements confirm that the effect of ISC on the emission rate is as low as \SI{0.1}{\%}, consistent with the literature knowledge that the ISC yield is extremely low for DBT in \textit{p}DCB at cryogenic temperatures\cite{Nicolet2007}. Second, we did not take into account any dephasing, as we reached the linewidth reported for DBT in \textit{p}DCB at $T<\SI{2}{K}$, consistent with the measured lifetimes of the excited state.\cite{Verhart2016,Zirkelbach2022} We have confirmed this by extracting the excited state lifetime from the same intensity autocorrelation measurements.

\begin{figure}[t!] 
\centering 
\includegraphics[width=0.45\textwidth]{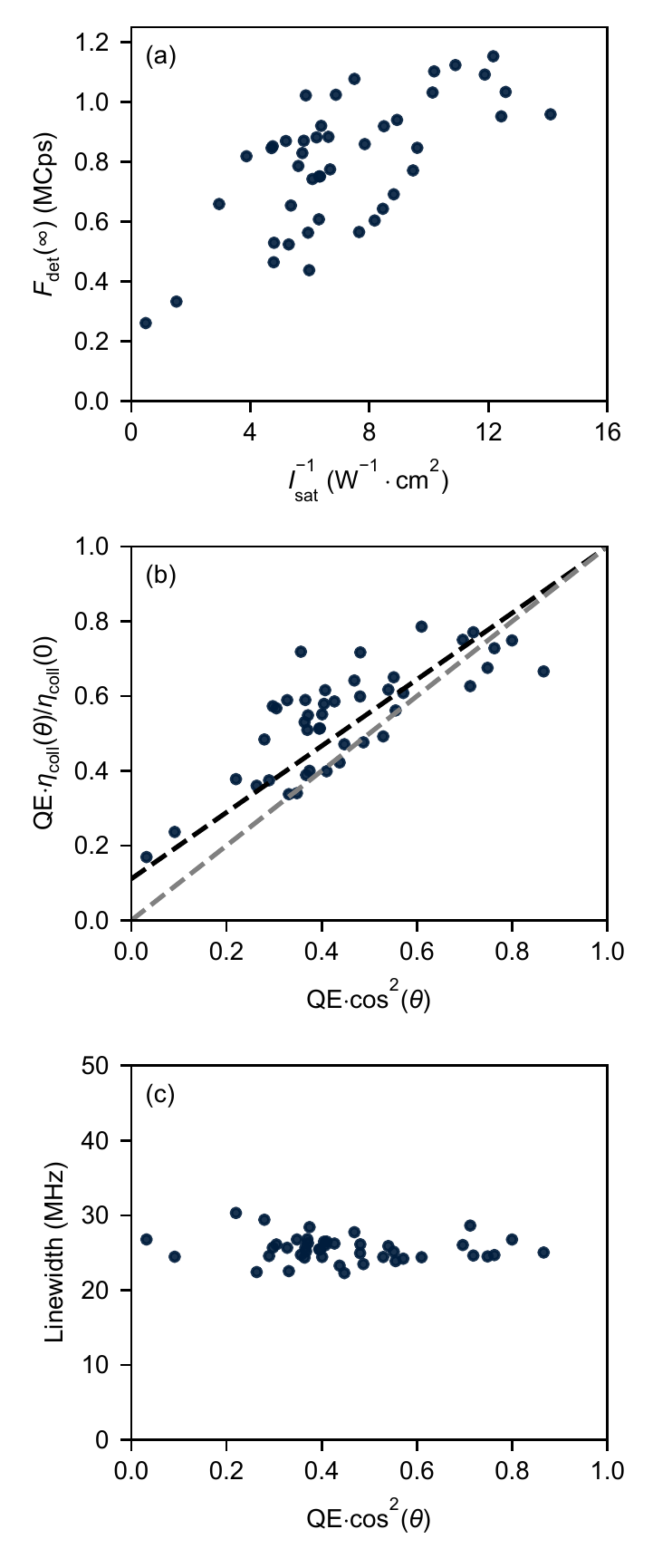}
\caption{(a) Joint distribution of the saturation fluorescence count rates and the inverse saturation power for the analyzed molecules. (b) Same, but for the extracted ``effective'' QE, which includes dipole orientation factors. The black and the grey dashed lines show the expected values assuming perfect $\mathrm{QE}=1$ with varying $0\leq\theta\leq\pi/2$ or perfect orientation $\theta=0$ with $0\leq\mathrm{QE}\leq 1$, respectively. (c) Joint distribution of the spectroscopic linewidth and the effective QE.} 
\label{fig:QE} 
\end{figure}

Figure \ref{fig:QE} summarizes the results obtained for 44 individual molecules. In Fig.\,\ref{fig:QE}(a), we show the recorded fluorescence rate as a function of the inverse saturation intensity $I_\mathrm{sat}^{-1}=A_\mathrm{eff}/P_\mathrm{sat}$. The values are well correlated, which is to be expected as both are equal to the product of the QE and a geometric $\theta$-dependent factor. Moreover, the results span over a large range of emission rates and saturation powers, indicating strong heterogeneity of the measured molecules. Although the QE of the individual molecules might differ, the observed strong heterogeneity likely stems from the variations in dipole orientation. 

It is instructive to examine the observations at the two ends of the range in Fig.\,\ref{fig:QE}(a). On the left side, there is a molecule with very high saturation intensity (more than 40 times above the median), which still produces relatively high saturation counts of \SI{0.25}{Mcps}. This is consistent with an almost perfectly out-of-plane dipole, which is hard to excite with transverse-polarized light, but whose fluorescence can still be collected with a finite efficiency $\eta_\mathrm{coll}\approx \SI{1}{\%}$. On the right side, there are molecules with a very high fluorescence count rate $\SI{>1.1}{Mcps}$ and correspondingly low saturation intensities, suggesting that a significant fraction of molecules has high QE.

Next, we use individually measured values of $\gamma_\mathrm{tot}$ and the knowledge of the setup calibration to examine the QE. As mentioned earlier, both methods rely on the dipole orientation. Therefore, in Fig.\,\ref{fig:QE}(b), we represent the results of the first and second methods via $\mathrm{QE}\cdot\eta_\mathrm{coll}(\theta)/\eta_\mathrm{coll}(0)$ and $\mathrm{QE}\cdot\cos^2(\theta)$, respectively. The two quantities are equal to the QE for the most favorable horizontal dipole ($\theta=0$) and underestimate the QE for other orientations. The best agreement between the two methods is obtained if we take $\alpha=0.33$ and assume that $\eta_\mathrm{spec}=0.8(1-\alpha)$, i.e., if the detection efficiency for the non-ZPL fluorescence is \SI{80}{\%} of that for ZPL contribution. This is reasonable considering the chromatic response of the collection and detection system over a spectral range of 100\,nm. A branching ratio of \SI{33}{\%} is also consistent with previously reported experimental values\cite{Verhart2016} and theoretical estimates\cite{ZirkelbachDiss} although a relative error of up to 50\% might be at play. 

The data points in Fig.\,\ref{fig:QE}(b) are spread over a wide range, while a nonnegligible fraction lies above \SI{70}{\%}. We estimate the overall uncertainty in our measurements to be about \SI{20}{\%}, which is substantially smaller than the variations observed in Fig.\,\ref{fig:QE}. Hence, we conclude that the QE of DBT in \textit{p}DCB may assume large values in the range of $70\pm 20\%$. In fact, in one single case (not shown in Fig.\,\ref{fig:QE}(b)), the extracted QE value significantly exceeded 1 for both methods used in this article. We suspect that the orientation and position of that molecule with respect to its surrounding matrix and substrate happened to inflict a planar antenna effect on it, leading to an increase in the excitation and collection efficiency\cite{Chu2017a}. We have excluded that molecule from our analysis.

Figure\,\ref{fig:QE}(c) shows that the measured distribution of the linewidths $\gamma_\mathrm{tot}$ is much narrower than the distribution of the data points presented in Fig.\,\ref{fig:QE}(b) and that there is no clear correlation between the two. Because we do not expect substantial variations of $\gamma_\mathrm{r}$ among the different molecules in the crystal, this observation can only be reconciled with the relation $\gamma_\mathrm{tot}=\gamma_\mathrm{r}/\mathrm{QE}$ via the distribution of the dipole orientation $\theta$. 

\section{Conclusions}

We have analyzed the quantum efficiency of DBT in \textit{p}DCB at liquid helium temperature using two different methods. The results agree well within reasonable assumptions about the branching ratio and the fluorescence collection efficiency, but in both cases we find a large spread in the extracted QE values. We attribute this distribution to the variation in the orientations of the individual molecules. Our analysis suggests that the QE of DBT reaches above \SI{70}{\%}. The discrepancy with the recently reported low QE values of DBT at ambient conditions \cite{Erker2022} can follow a number of reasons, one of them being a significant temperature dependence of the QE. Future studies should pursue comprehensive measurements using the same system under different conditions. Furthermore, quantification of the dipole orientation will be invaluable for more precise measurements, e.g., by employing high-NA optics or planar optical antennas \cite{Chu2017a} for collecting all polarizations. Finally, uncertainties in the branching ratio and host matrix birefringence and refractive index should be characterized more accurately.

This work was supported by the Max Planck Society and yjr Deutsche Forschungsgemeinschaft (DFG, German Research Foundation) – Project-ID 429529648 – TRR 306 QuCoLiMa (``Quantum Cooperativity of Light and Matter''). We are grateful to Prof. Thomas Basch\'{e} for the generous supply of purified DBT. We thank Luis Morales Inostroza for his assistance with the collection efficiency simulations.

\bibliography{DBT_QE}

\providecommand{\latin}[1]{#1}
\makeatletter
\providecommand{\doi}
  {\begingroup\let\do\@makeother\dospecials
  \catcode`\{=1 \catcode`\}=2 \doi@aux}
\providecommand{\doi@aux}[1]{\endgroup\texttt{#1}}
\makeatother
\providecommand*\mcitethebibliography{\thebibliography}
\csname @ifundefined\endcsname{endmcitethebibliography}
  {\let\endmcitethebibliography\endthebibliography}{}
\begin{mcitethebibliography}{43}
\providecommand*\natexlab[1]{#1}
\providecommand*\mciteSetBstSublistMode[1]{}
\providecommand*\mciteSetBstMaxWidthForm[2]{}
\providecommand*\mciteBstWouldAddEndPuncttrue
  {\def\EndOfBibitem{\unskip.}}
\providecommand*\mciteBstWouldAddEndPunctfalse
  {\let\EndOfBibitem\relax}
\providecommand*\mciteSetBstMidEndSepPunct[3]{}
\providecommand*\mciteSetBstSublistLabelBeginEnd[3]{}
\providecommand*\EndOfBibitem{}
\mciteSetBstSublistMode{f}
\mciteSetBstMaxWidthForm{subitem}{(\alph{mcitesubitemcount})}
\mciteSetBstSublistLabelBeginEnd
  {\mcitemaxwidthsubitemform\space}
  {\relax}
  {\relax}

\bibitem[Lakowicz(2006)]{Lakowicz2006principles}
Lakowicz,~J.~R. \emph{{Principles of fluorescence spectroscopy}}; Springer,
  2006\relax
\mciteBstWouldAddEndPuncttrue
\mciteSetBstMidEndSepPunct{\mcitedefaultmidpunct}
{\mcitedefaultendpunct}{\mcitedefaultseppunct}\relax
\EndOfBibitem
\bibitem[Siegman(1986)]{Siegman1986lasers}
Siegman,~A.~E. \emph{{Lasers}}; University science books, 1986\relax
\mciteBstWouldAddEndPuncttrue
\mciteSetBstMidEndSepPunct{\mcitedefaultmidpunct}
{\mcitedefaultendpunct}{\mcitedefaultseppunct}\relax
\EndOfBibitem
\bibitem[Johansen \latin{et~al.}(2008)Johansen, Stobbe, Nikolaev, Lund-Hansen,
  Kristensen, Hvam, Vos, and Lodahl]{Johansen2008}
Johansen,~J.; Stobbe,~S.; Nikolaev,~I.~S.; Lund-Hansen,~T.; Kristensen,~P.~T.;
  Hvam,~J.~M.; Vos,~W.~L.; Lodahl,~P. {Size dependence of the wavefunction of
  self-assembled InAs quantum dots from time-resolved optical measurements}.
  \emph{Physical Review B} \textbf{2008}, \emph{77}, 073303\relax
\mciteBstWouldAddEndPuncttrue
\mciteSetBstMidEndSepPunct{\mcitedefaultmidpunct}
{\mcitedefaultendpunct}{\mcitedefaultseppunct}\relax
\EndOfBibitem
\bibitem[Olutas \latin{et~al.}(2015)Olutas, Guzelturk, Kelestemur, Yeltik,
  Delikanli, and Demir]{Olutas2015}
Olutas,~M.; Guzelturk,~B.; Kelestemur,~Y.; Yeltik,~A.; Delikanli,~S.;
  Demir,~H.~V. {Lateral Size-Dependent Spontaneous and Stimulated Emission
  Properties in Colloidal CdSe Nanoplatelets}. \emph{ACS Nano} \textbf{2015},
  \emph{9}, 5041--5050\relax
\mciteBstWouldAddEndPuncttrue
\mciteSetBstMidEndSepPunct{\mcitedefaultmidpunct}
{\mcitedefaultendpunct}{\mcitedefaultseppunct}\relax
\EndOfBibitem
\bibitem[W{\"{u}}rth \latin{et~al.}(2012)W{\"{u}}rth, Gonz{\'{a}}lez, Niessner,
  Panne, Haisch, and Genger]{Wurth2012}
W{\"{u}}rth,~C.; Gonz{\'{a}}lez,~M.~G.; Niessner,~R.; Panne,~U.; Haisch,~C.;
  Genger,~U.~R. {Determination of the absolute fluorescence quantum yield of
  rhodamine 6G with optical and photoacoustic methods - Providing the basis for
  fluorescence quantum yield standards}. \emph{Talanta} \textbf{2012},
  \emph{90}, 30--37\relax
\mciteBstWouldAddEndPuncttrue
\mciteSetBstMidEndSepPunct{\mcitedefaultmidpunct}
{\mcitedefaultendpunct}{\mcitedefaultseppunct}\relax
\EndOfBibitem
\bibitem[Cranfill \latin{et~al.}(2016)Cranfill, Sell, Baird, Allen, Lavagnino,
  de~Gruiter, Kremers, Davidson, Ustione, and Piston]{Cranfill2016}
Cranfill,~P.~J.; Sell,~B.~R.; Baird,~M.~A.; Allen,~J.~R.; Lavagnino,~Z.;
  de~Gruiter,~H.~M.; Kremers,~G.-J.; Davidson,~M.~W.; Ustione,~A.;
  Piston,~D.~W. {Quantitative assessment of fluorescent proteins}. \emph{Nature
  Methods} \textbf{2016}, \emph{13}, 557--562\relax
\mciteBstWouldAddEndPuncttrue
\mciteSetBstMidEndSepPunct{\mcitedefaultmidpunct}
{\mcitedefaultendpunct}{\mcitedefaultseppunct}\relax
\EndOfBibitem
\bibitem[W{\"{u}}rth \latin{et~al.}(2015)W{\"{u}}rth, Gei{\ss}ler, Behnke,
  Kaiser, and Resch-Genger]{Wurth2015}
W{\"{u}}rth,~C.; Gei{\ss}ler,~D.; Behnke,~T.; Kaiser,~M.; Resch-Genger,~U.
  {Critical review of the determination of photoluminescence quantum yields of
  luminescent reporters}. \emph{Analytical and Bioanalytical Chemistry}
  \textbf{2015}, \emph{407}, 59--78\relax
\mciteBstWouldAddEndPuncttrue
\mciteSetBstMidEndSepPunct{\mcitedefaultmidpunct}
{\mcitedefaultendpunct}{\mcitedefaultseppunct}\relax
\EndOfBibitem
\bibitem[Orfield \latin{et~al.}(2016)Orfield, McBride, Wang, Buck, Keene, Reid,
  Htoon, Hollingsworth, and Rosenthal]{Orfield2016}
Orfield,~N.~J.; McBride,~J.~R.; Wang,~F.; Buck,~M.~R.; Keene,~J.~D.;
  Reid,~K.~R.; Htoon,~H.; Hollingsworth,~J.~A.; Rosenthal,~S.~J. {Quantum Yield
  Heterogeneity among Single Nonblinking Quantum Dots Revealed by Atomic
  Structure-Quantum Optics Correlation}. \emph{ACS Nano} \textbf{2016},
  \emph{10}, 1960--1968\relax
\mciteBstWouldAddEndPuncttrue
\mciteSetBstMidEndSepPunct{\mcitedefaultmidpunct}
{\mcitedefaultendpunct}{\mcitedefaultseppunct}\relax
\EndOfBibitem
\bibitem[Mohtashami and {Femius Koenderink}(2013)Mohtashami, and {Femius
  Koenderink}]{Mohtashami2013}
Mohtashami,~A.; {Femius Koenderink},~A. {Suitability of nanodiamond
  nitrogen-vacancy centers for spontaneous emission control experiments}.
  \emph{New Journal of Physics} \textbf{2013}, \emph{15}, 043017\relax
\mciteBstWouldAddEndPuncttrue
\mciteSetBstMidEndSepPunct{\mcitedefaultmidpunct}
{\mcitedefaultendpunct}{\mcitedefaultseppunct}\relax
\EndOfBibitem
\bibitem[Chizhik \latin{et~al.}(2011)Chizhik, Chizhik, Khoptyar, B{\"{a}}r,
  Meixner, and Enderlein]{Chizhik2011}
Chizhik,~A.~I.; Chizhik,~A.~M.; Khoptyar,~D.; B{\"{a}}r,~S.; Meixner,~A.~J.;
  Enderlein,~J. {Probing the radiative transition of single molecules with a
  tunable microresonator}. \emph{Nano Letters} \textbf{2011}, \emph{11},
  1700--1703\relax
\mciteBstWouldAddEndPuncttrue
\mciteSetBstMidEndSepPunct{\mcitedefaultmidpunct}
{\mcitedefaultendpunct}{\mcitedefaultseppunct}\relax
\EndOfBibitem
\bibitem[Chu \latin{et~al.}(2017)Chu, G{\"o}tzinger, and Sandoghdar]{Chu2017a}
Chu,~X.-L.; G{\"o}tzinger,~S.; Sandoghdar,~V. {A single molecule as a
  high-fidelity photon gun for producing intensity-squeezed light}.
  \emph{Nature Photonics} \textbf{2017}, \emph{11}, 58--62\relax
\mciteBstWouldAddEndPuncttrue
\mciteSetBstMidEndSepPunct{\mcitedefaultmidpunct}
{\mcitedefaultendpunct}{\mcitedefaultseppunct}\relax
\EndOfBibitem
\bibitem[Kwadrin and Koenderink(2012)Kwadrin, and Koenderink]{Kwadrin2012}
Kwadrin,~A.; Koenderink,~A.~F. {Gray-Tone Lithography Implementation of
  Drexhage's Method for Calibrating Radiative and Nonradiative Decay Constants
  of Fluorophores}. \emph{Journal of Physical Chemistry C} \textbf{2012},
  \emph{116}, 16666--16673\relax
\mciteBstWouldAddEndPuncttrue
\mciteSetBstMidEndSepPunct{\mcitedefaultmidpunct}
{\mcitedefaultendpunct}{\mcitedefaultseppunct}\relax
\EndOfBibitem
\bibitem[Sandoghdar(2020)]{Sandoghdar-NL2020}
Sandoghdar,~V. {Nano-Optics in 2020 $\pm$ 20}. \emph{Nano Letters}
  \textbf{2020}, \emph{20}, 4721--4723\relax
\mciteBstWouldAddEndPuncttrue
\mciteSetBstMidEndSepPunct{\mcitedefaultmidpunct}
{\mcitedefaultendpunct}{\mcitedefaultseppunct}\relax
\EndOfBibitem
\bibitem[Drexhage(1974)]{Drexhage1974}
Drexhage,~K.~H. \emph{{IV Interaction of Light with Monomolecular Dye Layers}};
  1974; pp 163--232\relax
\mciteBstWouldAddEndPuncttrue
\mciteSetBstMidEndSepPunct{\mcitedefaultmidpunct}
{\mcitedefaultendpunct}{\mcitedefaultseppunct}\relax
\EndOfBibitem
\bibitem[Brokmann \latin{et~al.}(2004)Brokmann, Coolen, Dahan, and
  Hermier]{Brokmann2004}
Brokmann,~X.; Coolen,~L.; Dahan,~M.; Hermier,~J.~P. {Measurement of the
  radiative and nonradiative decay rates of single CdSe nanocrystals through a
  controlled modification of their spontaneous emission}. \emph{Physical Review
  Letters} \textbf{2004}, \emph{93}, 107403\relax
\mciteBstWouldAddEndPuncttrue
\mciteSetBstMidEndSepPunct{\mcitedefaultmidpunct}
{\mcitedefaultendpunct}{\mcitedefaultseppunct}\relax
\EndOfBibitem
\bibitem[Buchler \latin{et~al.}(2005)Buchler, Kalkbrenner, Hettich, and
  Sandoghdar]{Buchler2005}
Buchler,~B.~C.; Kalkbrenner,~T.; Hettich,~C.; Sandoghdar,~V. {Measuring the
  Quantum Efficiency of the Optical Emission of Single Radiating Dipoles Using
  a Scanning Mirror}. \emph{Physical Review Letters} \textbf{2005}, \emph{95},
  063003\relax
\mciteBstWouldAddEndPuncttrue
\mciteSetBstMidEndSepPunct{\mcitedefaultmidpunct}
{\mcitedefaultendpunct}{\mcitedefaultseppunct}\relax
\EndOfBibitem
\bibitem[Jelezko \latin{et~al.}(1996)Jelezko, Tamarat, Lounis, and
  Orrit]{Jelezko1996}
Jelezko,~F.; Tamarat,~P.; Lounis,~B.; Orrit,~M. {Dibenzoterrylene in
  Naphthalene: A New Crystalline System for Single Molecule Spectroscopy in the
  Near Infrared}. \emph{Journal of Physical Chemistry} \textbf{1996},
  \emph{100}, 13892--13894\relax
\mciteBstWouldAddEndPuncttrue
\mciteSetBstMidEndSepPunct{\mcitedefaultmidpunct}
{\mcitedefaultendpunct}{\mcitedefaultseppunct}\relax
\EndOfBibitem
\bibitem[Toninelli \latin{et~al.}(2021)Toninelli, Gerhardt, Clark,
  Reserbat-Plantey, G{\"o}tzinger, Ristanovi{\'{c}}, Colautti, Lombardi, Major,
  Deperasi{\'{n}}ska, Pernice, Koppens, Kozankiewicz, Gourdon, Sandoghdar, and
  Orrit]{Toninelli2021}
Toninelli,~C.; Gerhardt,~I.; Clark,~A.~S.; Reserbat-Plantey,~A.;
  G{\"o}tzinger,~S.; Ristanovi{\'{c}},~Z.; Colautti,~M.; Lombardi,~P.;
  Major,~K.~D.; Deperasi{\'{n}}ska,~I. \latin{et~al.}  {Single organic
  molecules for photonic quantum technologies}. \emph{Nature Materials}
  \textbf{2021}, \emph{20}, 1615--1628\relax
\mciteBstWouldAddEndPuncttrue
\mciteSetBstMidEndSepPunct{\mcitedefaultmidpunct}
{\mcitedefaultendpunct}{\mcitedefaultseppunct}\relax
\EndOfBibitem
\bibitem[Wang \latin{et~al.}(2019)Wang, Kelkar, Martin-Cano, Rattenbacher,
  Shkarin, Utikal, G{\"o}tzinger, and Sandoghdar]{Wang2019}
Wang,~D.; Kelkar,~H.; Martin-Cano,~D.; Rattenbacher,~D.; Shkarin,~A.;
  Utikal,~T.; G{\"o}tzinger,~S.; Sandoghdar,~V. {Turning a molecule into a
  coherent two-level quantum system}. \emph{Nature Physics} \textbf{2019},
  \emph{15}, 483--489\relax
\mciteBstWouldAddEndPuncttrue
\mciteSetBstMidEndSepPunct{\mcitedefaultmidpunct}
{\mcitedefaultendpunct}{\mcitedefaultseppunct}\relax
\EndOfBibitem
\bibitem[Pscherer \latin{et~al.}(2021)Pscherer, Meierhofer, Wang, Kelkar,
  Martin-Cano, Utikal, G{\"o}tzinger, and Sandoghdar]{Pscherer2021}
Pscherer,~A.; Meierhofer,~M.; Wang,~D.; Kelkar,~H.; Martin-Cano,~D.;
  Utikal,~T.; G{\"o}tzinger,~S.; Sandoghdar,~V. {Single-Molecule Vacuum Rabi
  Splitting: Four-Wave Mixing and Optical Switching at the Single-Photon
  Level}. \emph{Physical Review Letters} \textbf{2021}, \emph{127},
  133603\relax
\mciteBstWouldAddEndPuncttrue
\mciteSetBstMidEndSepPunct{\mcitedefaultmidpunct}
{\mcitedefaultendpunct}{\mcitedefaultseppunct}\relax
\EndOfBibitem
\bibitem[Trebbia \latin{et~al.}(2022)Trebbia, Deplano, Tamarat, and
  Lounis]{Trebbia2022}
Trebbia,~J.-B.; Deplano,~Q.; Tamarat,~P.; Lounis,~B. {Tailoring the
  superradiant and subradiant nature of two coherently coupled quantum
  emitters}. \emph{Nature Communications} \textbf{2022}, \emph{13}, 2962\relax
\mciteBstWouldAddEndPuncttrue
\mciteSetBstMidEndSepPunct{\mcitedefaultmidpunct}
{\mcitedefaultendpunct}{\mcitedefaultseppunct}\relax
\EndOfBibitem
\bibitem[Wrigge \latin{et~al.}(2008)Wrigge, Gerhardt, Hwang, Zumofen, and
  Sandoghdar]{Wrigge2008}
Wrigge,~G.; Gerhardt,~I.; Hwang,~J.; Zumofen,~G.; Sandoghdar,~V. {Efficient
  coupling of photons to a single molecule and the observation of its resonance
  fluorescence}. \emph{Nature Physics} \textbf{2008}, \emph{4}, 60--66\relax
\mciteBstWouldAddEndPuncttrue
\mciteSetBstMidEndSepPunct{\mcitedefaultmidpunct}
{\mcitedefaultendpunct}{\mcitedefaultseppunct}\relax
\EndOfBibitem
\bibitem[de~Vries and Wiersma(1979)de~Vries, and Wiersma]{DeVries1979}
de~Vries,~H.; Wiersma,~D.~A. {Fluorescence transient and optical free induction
  decay spectroscopy of pentacene in mixed crystals at 2 K. Determination of
  intersystem crossing and internal conversion rates}. \emph{Journal of
  Chemical Physics} \textbf{1979}, \emph{70}, 5807--5822\relax
\mciteBstWouldAddEndPuncttrue
\mciteSetBstMidEndSepPunct{\mcitedefaultmidpunct}
{\mcitedefaultendpunct}{\mcitedefaultseppunct}\relax
\EndOfBibitem
\bibitem[Erker and Basch\'{e}(2022)Erker, and Basch\'{e}]{Erker2022}
Erker,~C.; Basch\'{e},~T. {The Energy Gap Law at Work: {E}mission Yield and
  Rate Fluctuations of Single {NIR} Emitters}. \emph{Journal of the American
  Chemical Society} \textbf{2022}, \emph{144}, 14053--14056\relax
\mciteBstWouldAddEndPuncttrue
\mciteSetBstMidEndSepPunct{\mcitedefaultmidpunct}
{\mcitedefaultendpunct}{\mcitedefaultseppunct}\relax
\EndOfBibitem
\bibitem[Englman and Jortner(1970)Englman, and Jortner]{Englman1970}
Englman,~R.; Jortner,~J. {The energy gap law for radiationless transitions in
  large molecules}. \emph{Molecular Physics} \textbf{1970}, \emph{18},
  145--164\relax
\mciteBstWouldAddEndPuncttrue
\mciteSetBstMidEndSepPunct{\mcitedefaultmidpunct}
{\mcitedefaultendpunct}{\mcitedefaultseppunct}\relax
\EndOfBibitem
\bibitem[Turro \latin{et~al.}(2010)Turro, Ramamurthy, and Scaiano]{Turro2010}
Turro,~N.~J.; Ramamurthy,~V.; Scaiano,~J. \emph{{Modern Molecular
  Photochemistry of Organic Molecules}}; University Science Books Mill Valley,
  2010\relax
\mciteBstWouldAddEndPuncttrue
\mciteSetBstMidEndSepPunct{\mcitedefaultmidpunct}
{\mcitedefaultendpunct}{\mcitedefaultseppunct}\relax
\EndOfBibitem
\bibitem[Jang(2021)]{Jang2021}
Jang,~S.~J. {A simple generalization of the energy gap law for nonradiative
  processes}. \emph{Journal of Chemical Physics} \textbf{2021}, \emph{155},
  164106\relax
\mciteBstWouldAddEndPuncttrue
\mciteSetBstMidEndSepPunct{\mcitedefaultmidpunct}
{\mcitedefaultendpunct}{\mcitedefaultseppunct}\relax
\EndOfBibitem
\bibitem[Hofmann \latin{et~al.}(2005)Hofmann, Nicolet, Kol'chenko, and
  Orrit]{Hofmann2005}
Hofmann,~C.; Nicolet,~A. A.~L.; Kol'chenko,~M.~A.; Orrit,~M. {Towards
  nanoprobes for conduction in molecular crystals: Dibenzoterrylene in
  anthracene crystals}. \emph{Chemical Physics} \textbf{2005}, \emph{318},
  1--6\relax
\mciteBstWouldAddEndPuncttrue
\mciteSetBstMidEndSepPunct{\mcitedefaultmidpunct}
{\mcitedefaultendpunct}{\mcitedefaultseppunct}\relax
\EndOfBibitem
\bibitem[Verhart \latin{et~al.}(2016)Verhart, M{\"u}ller, and
  Orrit]{Verhart2016}
Verhart,~N.~R.; M{\"u}ller,~M.; Orrit,~M. {Spectroscopy of Single
  Dibenzoterrylene Molecules in para-Dichlorobenzene}. \emph{ChemPhysChem}
  \textbf{2016}, \emph{17}, 1524--1529\relax
\mciteBstWouldAddEndPuncttrue
\mciteSetBstMidEndSepPunct{\mcitedefaultmidpunct}
{\mcitedefaultendpunct}{\mcitedefaultseppunct}\relax
\EndOfBibitem
\bibitem[Nicolet \latin{et~al.}(2007)Nicolet, Hofmann, Kol'chenko,
  Kozankiewicz, and Orrit]{Nicolet2007}
Nicolet,~A. A.~L.; Hofmann,~C.; Kol'chenko,~M.~A.; Kozankiewicz,~B.; Orrit,~M.
  {Single Dibenzoterrylene Molecules in an Anthracene Crystal: Spectroscopy and
  Photophysics}. \emph{ChemPhysChem} \textbf{2007}, \emph{8}, 1215--1220\relax
\mciteBstWouldAddEndPuncttrue
\mciteSetBstMidEndSepPunct{\mcitedefaultmidpunct}
{\mcitedefaultendpunct}{\mcitedefaultseppunct}\relax
\EndOfBibitem
\bibitem[Rattenbacher \latin{et~al.}(2019)Rattenbacher, Shkarin, Renger,
  Utikal, G{\"o}tzinger, and Sandoghdar]{Rattenbacher2019}
Rattenbacher,~D.; Shkarin,~A.; Renger,~J.; Utikal,~T.; G{\"o}tzinger,~S.;
  Sandoghdar,~V. {Coherent coupling of single molecules to on-chip ring
  resonators}. \emph{New Journal of Physics} \textbf{2019}, \emph{21},
  062002\relax
\mciteBstWouldAddEndPuncttrue
\mciteSetBstMidEndSepPunct{\mcitedefaultmidpunct}
{\mcitedefaultendpunct}{\mcitedefaultseppunct}\relax
\EndOfBibitem
\bibitem[Walls and Milburn(2008)Walls, and Milburn]{Walls_milburn_2008}
Walls,~D.~F.; Milburn,~G.~J. \emph{{Quantum Optics}}, 2nd ed.; Springer-Verlag:
  Berlin, Heidelberg, 2008\relax
\mciteBstWouldAddEndPuncttrue
\mciteSetBstMidEndSepPunct{\mcitedefaultmidpunct}
{\mcitedefaultendpunct}{\mcitedefaultseppunct}\relax
\EndOfBibitem
\bibitem[Scully and Zubairy(1997)Scully, and Zubairy]{Scully}
Scully,~M.~O.; Zubairy,~M. \emph{{Quantum Optics}}; Cambridge University Press,
  1997\relax
\mciteBstWouldAddEndPuncttrue
\mciteSetBstMidEndSepPunct{\mcitedefaultmidpunct}
{\mcitedefaultendpunct}{\mcitedefaultseppunct}\relax
\EndOfBibitem
\bibitem[Tannoudlji \latin{et~al.}(2004)Tannoudlji, Dupont-Roc, and
  Crynberg]{Tannoudji}
Tannoudlji,~C.~C.; Dupont-Roc,~J.; Crynberg,~G. \emph{{Atom-Photons
  Interactions}}; Wiley-VCH, 2004\relax
\mciteBstWouldAddEndPuncttrue
\mciteSetBstMidEndSepPunct{\mcitedefaultmidpunct}
{\mcitedefaultendpunct}{\mcitedefaultseppunct}\relax
\EndOfBibitem
\bibitem[Toninelli \latin{et~al.}(2010)Toninelli, Early, Bremi, Renn,
  G{\"o}tzinger, and Sandoghdar]{Toninelli2010}
Toninelli,~C.; Early,~K.; Bremi,~J.; Renn,~A.; G{\"o}tzinger,~S.;
  Sandoghdar,~V. {Near-infrared single-photons from aligned molecules in
  ultrathin crystalline films at room temperature}. \emph{Optics Express}
  \textbf{2010}, \emph{18}, 6577\relax
\mciteBstWouldAddEndPuncttrue
\mciteSetBstMidEndSepPunct{\mcitedefaultmidpunct}
{\mcitedefaultendpunct}{\mcitedefaultseppunct}\relax
\EndOfBibitem
\bibitem[Pfab \latin{et~al.}(2004)Pfab, Zimmermann, Hettich, Gerhardt, Renn,
  and Sandoghdar]{Pfab2004}
Pfab,~R.~J.; Zimmermann,~J.; Hettich,~C.; Gerhardt,~I.; Renn,~A.;
  Sandoghdar,~V. {Aligned terrylene molecules in a spin-coated ultrathin
  crystalline film of p-terphenyl}. \emph{Chemical Physics Letters}
  \textbf{2004}, \emph{387}, 490--495\relax
\mciteBstWouldAddEndPuncttrue
\mciteSetBstMidEndSepPunct{\mcitedefaultmidpunct}
{\mcitedefaultendpunct}{\mcitedefaultseppunct}\relax
\EndOfBibitem
\bibitem[Lieb \latin{et~al.}(2004)Lieb, Zavislan, and Novotny]{Lieb2004}
Lieb,~M.~A.; Zavislan,~J.~M.; Novotny,~L. {Single-molecule orientations
  determined by direct emission pattern imaging}. \emph{J. Opt. Soc. Am. B}
  \textbf{2004}, \emph{21}, 1210--1215\relax
\mciteBstWouldAddEndPuncttrue
\mciteSetBstMidEndSepPunct{\mcitedefaultmidpunct}
{\mcitedefaultendpunct}{\mcitedefaultseppunct}\relax
\EndOfBibitem
\bibitem[Manghi \latin{et~al.}(1967)Manghi, de~Caroni, de~Benyacar, and
  de~Abeledo]{Manghi1967}
Manghi,~E.; de~Caroni,~C.~A.; de~Benyacar,~M.~R.; de~Abeledo,~M.~J. {Optical
  properties of p -dichlorobenzene}. \emph{Acta Crystallographica}
  \textbf{1967}, \emph{23}, 205--208\relax
\mciteBstWouldAddEndPuncttrue
\mciteSetBstMidEndSepPunct{\mcitedefaultmidpunct}
{\mcitedefaultendpunct}{\mcitedefaultseppunct}\relax
\EndOfBibitem
\bibitem[Bernard \latin{et~al.}(1993)Bernard, Fleury, Talon, and
  Orrit]{Bernard1993}
Bernard,~J.; Fleury,~L.; Talon,~H.; Orrit,~M. {Photon bunching in the
  fluorescence from single molecules: A probe for intersystem crossing}.
  \emph{The Journal of Chemical Physics} \textbf{1993}, \emph{98},
  850--859\relax
\mciteBstWouldAddEndPuncttrue
\mciteSetBstMidEndSepPunct{\mcitedefaultmidpunct}
{\mcitedefaultendpunct}{\mcitedefaultseppunct}\relax
\EndOfBibitem
\bibitem[Basch\'{e} \latin{et~al.}(1997)Basch\'{e}, Moerner, Orrit, and
  Wild]{Basche2008}
Basch\'{e},~T.; Moerner,~W.~E.; Orrit,~M.; Wild,~U.~P. \emph{{Single-Molecule
  Optical Detection, Imaging and Spectroscopy}}; Wiley-VCH, 1997\relax
\mciteBstWouldAddEndPuncttrue
\mciteSetBstMidEndSepPunct{\mcitedefaultmidpunct}
{\mcitedefaultendpunct}{\mcitedefaultseppunct}\relax
\EndOfBibitem
\bibitem[Zirkelbach \latin{et~al.}(2022)Zirkelbach, Mirzaei,
  Deperasi{\'{n}}ska, Kozankiewicz, Gurlek, Shkarin, Utikal, G{\"o}tzinger, and
  Sandoghdar]{Zirkelbach2022}
Zirkelbach,~J.; Mirzaei,~M.; Deperasi{\'{n}}ska,~I.; Kozankiewicz,~B.;
  Gurlek,~B.; Shkarin,~A.; Utikal,~T.; G{\"o}tzinger,~S.; Sandoghdar,~V.
  {High-resolution vibronic spectroscopy of a single molecule embedded in a
  crystal}. \emph{Journal of Chemical Physics} \textbf{2022}, \emph{156},
  104301\relax
\mciteBstWouldAddEndPuncttrue
\mciteSetBstMidEndSepPunct{\mcitedefaultmidpunct}
{\mcitedefaultendpunct}{\mcitedefaultseppunct}\relax
\EndOfBibitem
\bibitem[Zirkelbach(2021)]{ZirkelbachDiss}
Zirkelbach,~J. {High-Resolution Spectroscopy of Vibronic Transitions in Single
  Molecules}. Doctoral thesis, Friedrich-Alexander-Universit{\"a}t
  Erlangen-N{\"u}rnberg, 2021\relax
\mciteBstWouldAddEndPuncttrue
\mciteSetBstMidEndSepPunct{\mcitedefaultmidpunct}
{\mcitedefaultendpunct}{\mcitedefaultseppunct}\relax
\EndOfBibitem
\end{mcitethebibliography}
\end{document}